\documentclass[referee,psfig]{aa}
\usepackage{graphicx}

\newcommand{\kms}   {~km~s$^{-1}$}

\newcommand{\jy}   {~Jy~beam$^{-1}$}

\newcommand{\vlsr}  {$V_{\rm LSR}$}

\newcommand{\mo}    {$M_{\sun}$}

\newcommand{\meta}  {CH$_3$OH}

\newcommand{\arcdeg}{\mbox{$^\circ$}} 
\newcommand{\juz}   {\mbox{($J$=1$\rightarrow$0)}}
\newcommand{\jcc}   {\mbox{($J$=5$\rightarrow$4)}}
\newcommand{\jdu}   {\mbox{($J$=2$\rightarrow$1)}}
\newcommand{\jcucu}   {\mbox{($J$=5$_{-1}\rightarrow$4$_{-1}$)}}
\newcommand{\jr}[2] {\mbox{$J$=#1$\rightarrow$#2}}
\newcommand{\lesssim}{\mathrel{\hbox{\rlap{\hbox{\lower4pt\hbox{$\sim$}}}\hbox{$<$}}}}
\newcommand{\iras} {IRAS~21391+5802}

\begin{document}


\title{The Dense Molecular Cores in the IRAS~21391+5802 region}

\author{
Maria T.\    Beltr\'an\inst{1, 2}, 
Josep M.       Girart\inst{3, 4},
Robert       Estalella\inst{3},
and Paul T.\ P.\ Ho\inst{2}
}

\offprints{Maria T.\ Beltr\'an} 

\institute{
Osservatorio Astrofisico di Arcetri, Largo E.\ Fermi, 5, I-50125 Firenze, Italy
\and
Harvard-Smithsonian Center for Astrophysics, 60 Garden Street, Cambridge,
MA 02138, USA
\and
Departament d'Astronomia i Meteorologia, Universitat de Barcelona,
Av.\ Diagonal 647, 08028 Barcelona, Catalunya, Spain
\and
Institut de Ci\`encies de l'Espai (CSIC)-IEEC, Gran Capit\`a 2,
08034 Barcelona, Catalunya, Spain
}

\date{Received , Accepted }

\authorrunning{Beltr\'an et al.}
\titlerunning{The Dense Molecular Cores in the IRAS~21391+5802 region}   


\abstract{ We present a detailed kinematical study and modeling of the emission
of the molecular cores at ambient velocities surrounding IRAS~21391+5802, an
intermediate-mass protostar embedded in IC~1396N. The high-density gas
emission  is found in association with three dense cores associated with the YSOs BIMA~1, BIMA~2, and BIMA~3. The CS \jcc\ and \meta\ \jcucu\ emission around
BIMA~1 has been modeled by considering a spatially infinitely thin ring seen edge-on by
the observer. From the model we find that CS is detected over a wider radii
range than \meta. A bipolar outflow is detected in the CS \jdu\ line centered near BIMA~1. This outflow could be powered by a yet undetected YSO, BIMA~1W, or alternatively could be part of the BIMA~1 molecular outflow. The CS and
\meta\ emission associated with the intermediate-mass protostar BIMA~2 is
highly perturbed by the bipolar outflow even at cloud velocities, confirming
that the  protostar is in a very active stage of mass loss. For YSO BIMA~3 
the lack of outflow and of clear evidence of infall suggests
that both outflow and infall are weaker than in BIMA 1, and that BIMA 3 is
probably a more evolved object.
  }   

\maketitle

\section{Introduction\label{intro}}

It is well known that Young Stellar Objects (YSOs) are embedded in dust and
gas. For low-mass stars, the youngest and more embedded Class~0 objects have
most of their emission in large scale gas and dust structures, consistent with
envelopes, while more evolved optically visible T~Tauri stars have compact
continuum emission, associated with disks, with little or no extended emission
(e.g.\ Ohashi et al.\ \cite{ohashi91}, \cite{ohashi96}; Chen et al.\
\cite{chen92}; Looney et al.\ \cite{looney00}). This well-defined
evolutionary sequence found for low-mass stars, however, does not have a
counterpart for higher mass stars, such as Herbig~Ae and Be stars and their
intermediate-mass precursors with masses in the range 2~\mo$\leq
M_\star\leq10$~\mo. For the intermediate-mass YSOs, the difficulty of
identifying the youngest protostars, which are still in the infalling envelope
phase, and the fact that they tend to be located  at greater distances make
them less studied and less well understood objects. The immediate vicinity of
such protostars is a very complex environment, where the extended emission is
usually resolved  into more than one source when observed at high resolution
(e.g.\ G173.58+2.45: Shepherd \& Watson \cite{shepherd02}; IRAS~21391+5802:
Beltr\'an et al.\ \cite{beltran02}, hereafter Paper I). In addition, the
molecular outflows driven by intermediate-mass objects are more energetic.
Thus, their interaction with the circumstellar gas and dust material
surrounding the  protostars is expected to be stronger and more dramatic,
disrupting the envelopes and pushing away the dense gas at high velocities.

For the intermediate-mass protostar IRAS~21391+5802 we conducted a
detailed high angular resolution study by carrying out centimeter and
millimeter continuum and spectral line BIMA observations (Paper I). The
continuum emission at centimeter and millimeter wavelengths was resolved  into
three sources: BIMA~1, BIMA~2, and  BIMA~3. The strongest source at millimeter
wavelengths is BIMA~2, which is most likely the object associated with
IRAS~21391+5802, and it has a circumstellar mass of $\sim5.1\,M_\odot$.
Regarding the CO emission, at least two molecular outflows were detected in the
region, one of them is elongated along the north-south direction centered on BIMA~1,  and the
other one is a stronger east-west outflow centered at the position of  the
intermediate-mass protostar BIMA~2. The BIMA~2 CO outflow presents a complex structure and kinematics.
While at high outflow velocities the outflow is clearly bipolar, at low outflow
velocities the blueshifted and redshifted emission are highly overlapping, and
the strongest emission shows a V-shaped morphology. The CS
and CH$_3$OH emission from the BIMA~2 outflow exhibits two well differentiated and clumpy lobes, with two
prominent northern blueshifted and redshifted clumps. The curved shape of the
clumps and the spectral shape at these positions are consistent with shocked
material. In addition, CS and CH$_3$OH are strongly enhanced toward these
positions with respect to typical quiescent material abundances in other
star-forming regions. These kinematical and chemical evidences suggest that the
clumps are tracing gas entrained within the surface of interaction between the
molecular outflow and the dense ambient quiescent core, and that the morphology
of the molecular outflow is a result of this interaction. As compared to the
low-mass counterparts, the properties of the outflow driven by BIMA~2 are
consistent with those of the outflows driven by low-mass young objects, and it
fits well the correlations found for low-mass outflows by Cabrit \& Bertout
(\cite{cabrit92}), Anglada (\cite{ang96}), and Bontemps et al.
(\cite{bontemps96}). In addition, the morphology and properties of the
circumstellar structures around this intermediate-mass protostar are similar to
those of low-mass counterpart Class~0 objects. The other two sources in the region BIMA~1 and BIMA~3, have a mass of $\sim0.07\,M_\odot$ each, and their small dust emissivity index is suggestive of
grain growth in dense regions (e.g.\ Mannings \& Emerson \cite{mannings94}). This fact
together with the more compact appearance of their dust emission suggest that
they could be more evolved low-mass objects. The results presented in Paper~I analyzed and discussed the kinematics of the dense gas as it relates to the molecular outflows detected in the region. On the other hand, in order to continue our study
of this intermediate-mass star-forming region, in this present work we analyze and model the kinematics of the gas at ambient velocities as it relates to the molecular cores in which BIMA~1, BIMA~2, and BIMA~3 are embedded. The results of this study are presented here.

\section{Observations}

The details of the
observations, carried out with the 10--antenna BIMA array\footnote{ The BIMA array is operated by the
Berkeley--Illinois--Maryland Association with support from the National Science
Foundation.} between 2000 March and 2001 January, are given in Paper I.  The maps at 3.1 mm were done by using a
robust weight of 0.5, with a resulting synthesized beam of
$7\farcs0\times6\farcs3$,  P.A.$=-21\arcdeg$. Channel maps at 1.2~mm were also
done with a robust weight of 0.5, with a resulting synthesized beam of
$2\farcs1\times1\farcs8$, P.A.$=-3\arcdeg$, and the integrated intensity maps
were done with a robust weight of 1, and the resulting synthesized beam is
$2\farcs7\times2\farcs4$, P.A.$=-16\arcdeg$. The spectral resolution of the
observations was 0.30\kms\ for CS \jdu\ at 97.981~GHz, and 0.12\kms\ for CS
\jcc\ at 244.936~GHz and \meta\ \jcucu\ at 241.767~GHz. We note that for the
velocity range at which the ambient gas arises (between $\sim -2$ and 3\kms)
the CS \jdu\ emission appears to be quite extended (see Fig.~1). This suggests
that there is possibly a significant fraction of the total flux missed by the
BIMA interferometer.  The CS \jdu\ BIMA observations were done in the C array,
so these observations should be sensitive to structures up to 60$''$ (Wright
1996).  The dense cores associated with BIMA~1, BIMA~2 and BIMA~3 have sizes
(FWHM $\sim 15''$) significantly  below this value. The 1~mm observations
(\meta\ and CS \jcc) were done using the B and C arrays, so they should be
sensitive to structures up to 24$''$.  This value is still larger than the CS 
\jdu\ sizes of the three dense cores. Therefore, the 3 and 1~mm line emission
from the three dense cores studied in this paper should not be significantly
affected by missing flux.

\section{Results}

\subsection{Lines observed}

%
     \begin{figure}
     \begin{center}
     \rotatebox{270}{
     \resizebox{9cm}{!}{\includegraphics{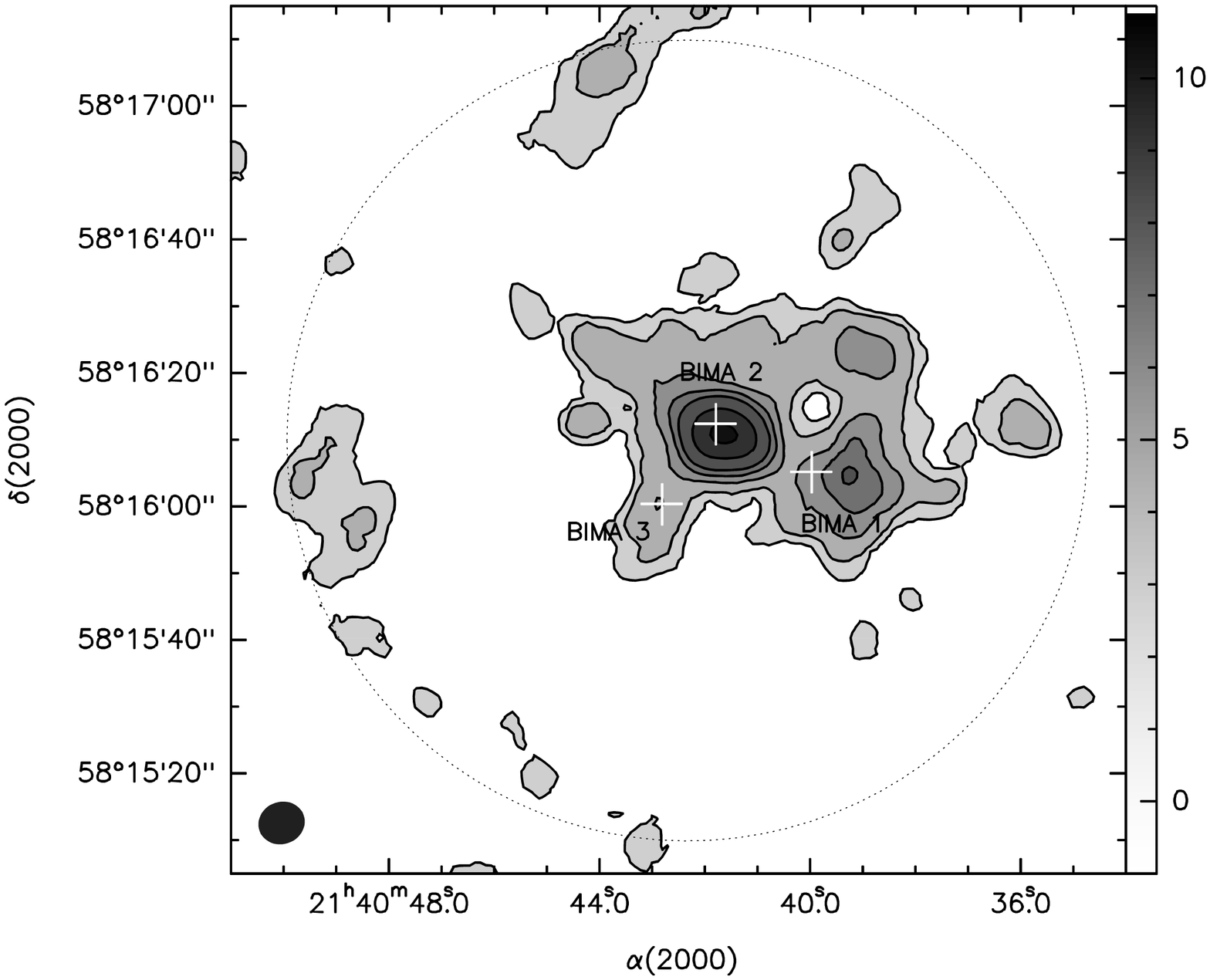}}}
     \hfill
     \caption[]{Integrated intensity of the CS \jdu\ emission over the
     velocity interval $(-3, 3)$\kms. The contours are  3, 5, 10, 15, 20,
     and 40 times 0.24\jy\kms. The map has not been corrected for primary beam response. The crosses show the position of the 3.1~mm
     sources BIMA 1, BIMA 2, and BIMA 3 (Paper I), in order of right ascension.
     The dotted circumference represents the BIMA primary beam (50\% attenuation
     level). The
     synthesized beam is shown in the lower left-hand corner.}
     \label{fcsto}
     \end{center}
     \end{figure}

CS \jdu, CS \jcc, and \meta\  \jcucu\ were observed toward \iras.
Figure~\ref{fcsto} shows the integrated emission of the CS \jdu\ at  ambient
velocities $(-3, 3)$\kms.  The strongest emission arises from a core associated
with the YSO BIMA~2.  West to this core, there is another core, which is associated
with the YSO BIMA~1.  Surrounding these two cores, there is extended and weaker
emission spanning about 20-30$''$ from the cores. The other YSO detected in the
region, BIMA~3, is also associated with CS emission. Additionally, an
elongation  is also visible toward the northeast of BIMA~2. This emission is
clearly seen in CS \jcc\ as a clump located $\sim 15''$ from BIMA~2 (see
Fig.~11 of Paper~I).

CS \jcc\ and \meta\ \jcucu\ emission are also associated with the three YSOs
BIMA~1, BIMA~2, and BIMA~3. The strongest and more extended emission for both
species arises from the core associated with BIMA~2. Figure~2 shows the spectra
obtained for CS \jdu, CS \jcc, and \meta\ \jcucu\ toward the positions of
BIMA~1, BIMA~2, and BIMA~3.



\subsection{Physical parameters
\label{phy-pa}}

%
     \begin{figure}
     \label{fspectre}
     \begin{center}
     \rotatebox{270}{
     \resizebox{10cm}{!}{\includegraphics{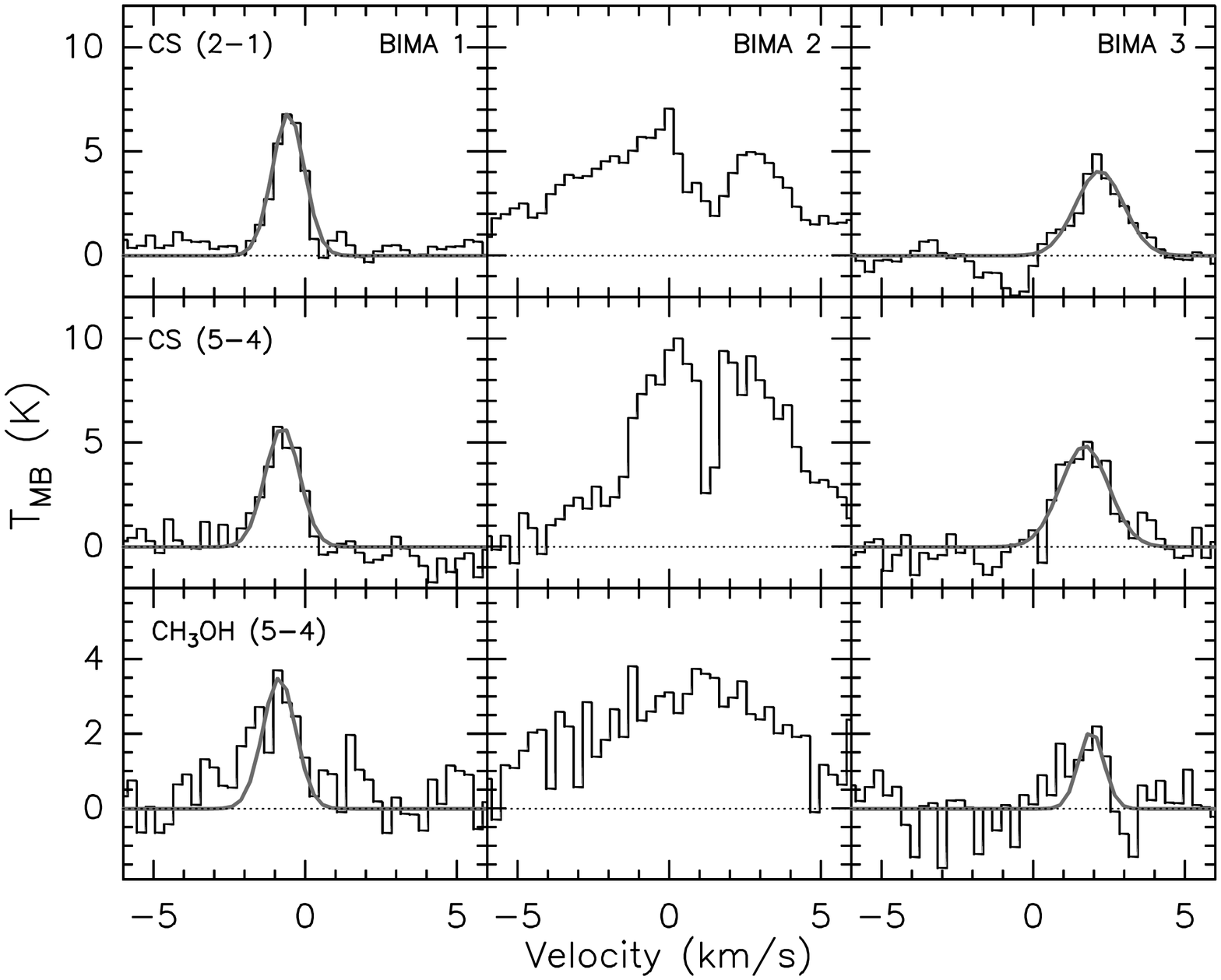}}}
     \hfill
     \caption[]{Spectra of the CS \jdu, CS \jcc\ and \meta\ \jcucu\ transitions
     toward the positions of BIMA~1, BIMA~2, and BIMA~3. The
     continuum emission has been subtracted. The 1-$\sigma$ noise in 1 channel
     is 0.3~K
     for CS \jdu, and 0.7~K for CS \jcc, and \meta\ \jcucu. The conversion factor is 2.92 K/Jy\,beam$^{-1}$ for CS \jdu, and 5.24 K/Jy\,beam$^{-1}$ for CS \jcc, and \meta\ \jcucu.  The thick grey profiles are the Gaussian fits to the spectra of BIMA~1 and BIMA~3.} 
     \end{center}
     \end{figure}
%

\begin{table}
\caption[]{CS (\jr{2}{1}), CS (\jr{5}{4}), and \meta\ \jcucu\ lines and physical
parameters obtained toward the positions of BIMA 1 and BIMA 3 around IRAS 21391-5802.}
     \label{tparam}
\begin{tabular}{llcccccc}
\hline\noalign{\smallskip}
 & Molecule \& & $\int T_{\rm mb}{\rm d}v$ &$T_{\rm mb}$ & $V_{\rm 
LSR}$ & $\Delta V$  &$\theta_{\rm CS}~^{(a)}$ & $M_{\rm vir}~^{(b)}$ \\
Position &Transition        & (K \kms) & (K) & (\kms) & (\kms) & (arcsec)
& (\mo)\\
\noalign{\smallskip}\hline\noalign{\medskip}
BIMA 1  &CS \jdu\      &  $9.1\pm0.5$ &   $6.9\pm0.5$ & $-0.54\pm0.03$
             & $1.25\pm0.08$  \\
        &CS \jcc\  & $8.2\pm0.7$ &   $5.8\pm0.7$
             &  $-0.74\pm0.05$ & $1.33\pm0.13$ &$3.2\pm0.2$ & $0.31\pm0.12$\\
        &\meta\  &   $4.8\pm0.8$ &   $3.5\pm0.7$
             & $-0.74\pm0.10$ & $1.27\pm0.27$\\ 
\noalign{\smallskip}\hline\noalign{\smallskip}
BIMA 3  &CS \jdu\ & $8.1\pm0.5$ &   $4.1\pm0.4$
             & $+2.21\pm0.05$ & $1.86\pm0.15$\\
          &CS \jcc\ &  $9.7\pm0.7$ &  $4.9\pm0.8$
             & $+1.71\pm0.07$ &$1.87\pm0.17$  &$3.4\pm0.2$  & ~~0.3 -- 4.6$^{(c)}$ \\
          &\meta\  &   $2.2\pm0.5$ &   $2.1\pm0.6$
             &$+1.93\pm0.13$ & $0.98\pm0.23$s \\
\noalign{\medskip}\hline\noalign{\medskip}
\end{tabular}
   {\footnotesize
\noindent $^{(a)}$ Deconvolved geometric mean of the major and minor axes of the
 50\% of the peak contour of the CS (\jr{5}{4}) map.  
        
\noindent $^{(b)}$ Virial mass obtained from the CS (\jr{5}{4}) using a $\Delta V$ of 0.5~\kms, intrinsic line width (i.e.\ the line width corrected for kinematics) obtained from the
collapsing ring model for
BIMA 1 (see $\S~\ref{modelbima1}$).

\noindent $^{(c)}$  We cannot disentangle the kinematics of BIMA 3.
Therefore we give the values for the virial mass assuming that the CS line width
is the intrinsic line width (which gives an upper limit), and the value adopting
$\Delta V=0.5$~\kms, the intrinsic line width derived for BIMA 1.}
 \end{table}

Table~\ref{tparam} lists the fitted line parameters for each transition for
BIMA~1 and BIMA~3. The core associated with the YSO BIMA~2 is being strongly
disturbed by the outflow even at velocities close to the systemic value,
$V_{\rm  LSR}\simeq 0$\kms. As can be seen in the spectra in Figure~2, 
the BIMA~2 profiles are not Gaussian and show emission at high velocities. This is probably because the molecular outflow dominates
the emission. In addition, both CS \jdu\  and \jcc\ lines also show a clear
redshifted self-absorption feature. Thus, because of the difficulty to
disentangle the cloud emission  from the outflow emission, we cannot study the
properties of the molecular core associated with BIMA~2. The outflow clearly
enhanced the integrated emission map shown in Figure~\ref{fcsto}.

From the spectra (Fig.~2) and their fits (Table~\ref{tparam}) it is clear that
while BIMA~1 has a systemic velocity of
$\sim -0.7$\kms, the core associated with BIMA~3 is at a different
velocity, $V_{\rm  LSR}\simeq 2$\kms. In Table~\ref{tparam} we also show the deconvolved size of the cores obtained
from the CS \jcc\ data, and the virial mass estimated from the intrinsic line width (i.e.\ the line width corrected for kinematics) obtained from the
collapsing ring model for
BIMA 1 (see $\S~\ref{modelbima1}$), assuming a homogeneous spherical clump,
and neglecting contributions from magnetic field and surface pressure. In such a
case the virial mass can be computed from the expression (see e.g.\ MacLaren et
al.\ \cite{maclaren88}): 
\begin{equation} 
\bigg( \frac{{M_{\rm vir}}}{M_\odot} \bigg) =
0.509 \bigg(\frac{d}{{\rm kpc}}\bigg) \bigg(\frac{\theta}{{\rm arcsec}}\bigg)
\bigg(\frac{\Delta V}{{\rm km~s}^{-1}}\bigg)^2 
\label{virial} 
\end{equation}
where $d$ is the distance, 750~pc for IRAS~21391+5802 (Matthews
\cite{matthews79}),
$\theta$ is the deconvolved size of the source, and
${\Delta V}$ is the intrinsic line width. It should be noted that the virial
mass depends on the density profile. For a power-law density distribution of
the type $\rho\propto r^{-p}$, the virial mass obtained from Eq.~(\ref{virial})
should be multiplied by a factor 
$3(5-2p)/5(3-p)$ 
which is $\leq 1$ for $p<3$ (MacLaren et
al.\ \cite{maclaren88}). Thus, the values given in Table~\ref{tparam}
should be taken as upper limits.

\section{Discussion}

\subsection{BIMA 1}

     \begin{figure}
     \begin{center}
     \rotatebox{270}{
     \resizebox{6.5cm}{!}{\includegraphics{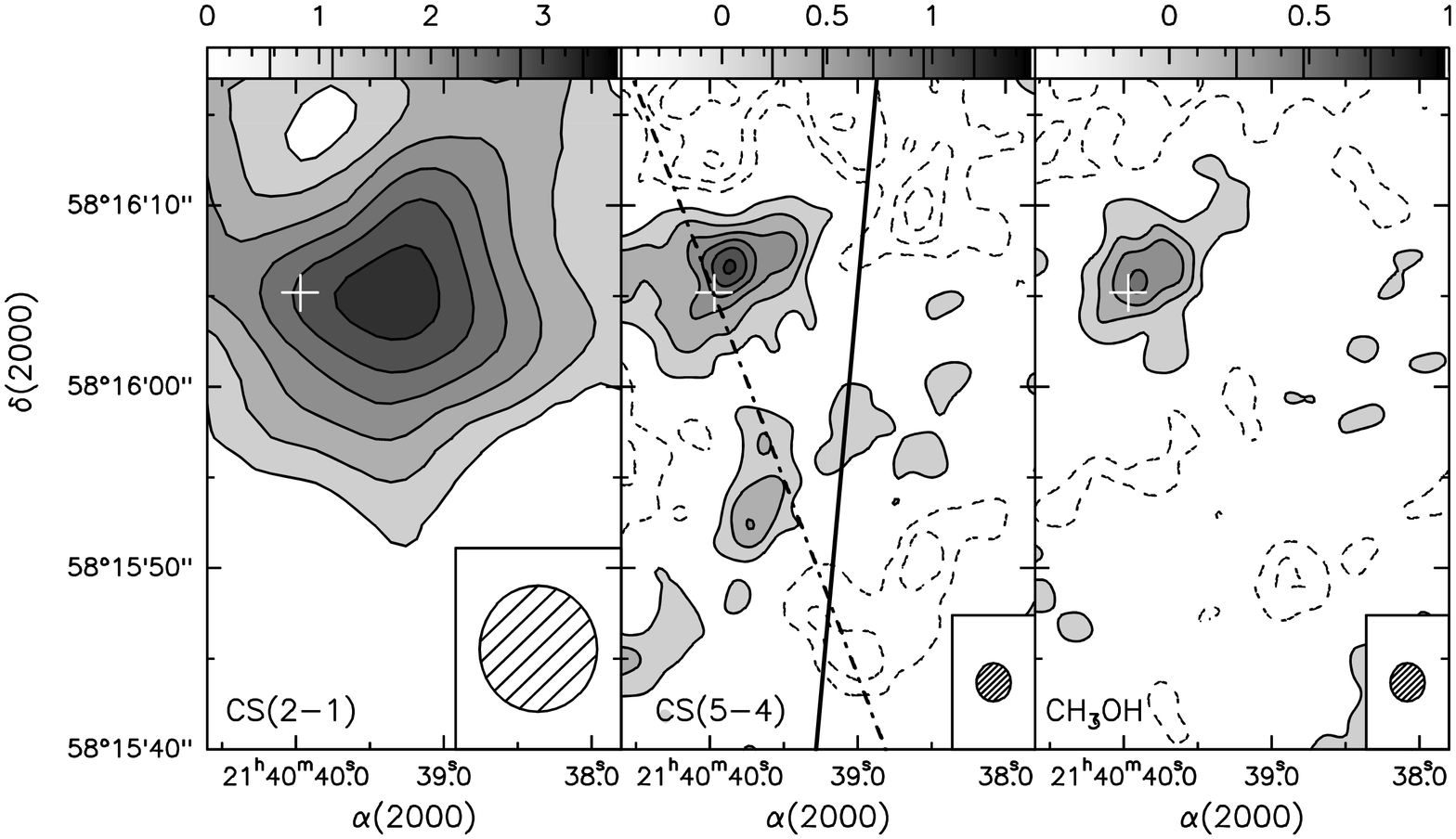}}}
     \hfill
     \caption[]{BIMA~1: Integrated intensity of the CS \jdu, CS \jcc, and \meta\
\jcucu\  emission over the velocity interval $(-1.8, 0.1)$\kms\ toward BIMA~1.
The contours are 0.56 to 3.36\jy\kms\ in steps of 0.56\jy\kms\ for CS \jdu, $-0.72$, $-0.48$,
$-0.24$, 0.24 to 1.44\jy\kms\ in steps of 0.24\jy\kms\ for CS \jcc, and $-0.48$,
$-0.24$, 0.24 to 0.96\jy\kms\ in steps of 0.24\jy\kms\ for \meta\ \jcucu. The cross shows the position of the 3.1~mm source BIMA 1 (Paper I). The synthesized beams are shown in the lower right-hand corner. The thick dashed-dotted line in the middle panel marks the axis of the north-south CO outflow driven by the YSO BIMA~1, at P.A.\ $\simeq 20\degr$ (Paper I). The thick solid line marks the axis of
the CS \jdu\ blue and red lobes, with a P.A.\ $\simeq -5\degr$ (see $\S$~\ref{bima1_cs}).}
     \label{fintbima1}
     \end{center}
     \end{figure}

As seen in the previous section the high-density gas emission at ambient
velocities $(-3, 3)$ \kms\ around IRAS~21391+5802 is found in association with
three dense cores associated with the YSOs BIMA~1, BIMA~2, and
BIMA~3. We conducted a
kinematical study of each core that revealed important properties of the cores
and of their evolutionary stage.   

Figure~\ref{fintbima1} shows the emission of CS \jdu, CS \jcc, and \meta\
\jcucu\ toward BIMA~1, integrated for the velocity range $(-1.8, 0.1)$ \kms.
The core associated with the YSO BIMA~1, which is likely the powering source of
the north-south bipolar molecular outflow detected in the region (Paper I), is
well traced by the three transitions. On the one hand, CS \jcc\ and \meta\ peak at the same position as the 3.1~mm continuum emission (Paper~I). On the other hand, as can be seen in this figure, the CS \jdu\ emission peaks at a position shifted $\sim 5''$ to the west from that of the CS \jcc\ and \meta\ emissions. This shift in position is of the
order of the CS \jdu\ beam ($\sim 7''$), and thus, significant. Note that this
is not an instrumental or calibration effect, since the observations were
carried out by configuring the BIMA digital correlator to observe
simultaneously the 3.1~mm continuum emission and the CS \jdu\ line. 

The morphology of the emission of the three species is also different. 
Despite the CS \jdu\ observations having a lower angular resolution than CS
\jcc, CS \jdu\  is tracing a more extended region, with a deconvolved size of
$\sim 14''$, about 3 times larger than that traced by CS \jcc. All this
suggests that CS \jdu\ is tracing different material than CS \jcc\ and \meta,
and thus, the physical properties and kinematics of the gas traced by them 
might be different. CS \jcc\ and \meta\  would be likely tracing dense
and hot gas associated with the YSO BIMA~1, while CS \jdu\ would be tracing more
extended, less dense and cooler material.

\subsubsection{CS \jcc\ and \meta\ \jcucu}
\label{modelbima1}

     \begin{figure}
     \begin{center}
     \rotatebox{270}{
     \resizebox{9cm}{!}{\includegraphics{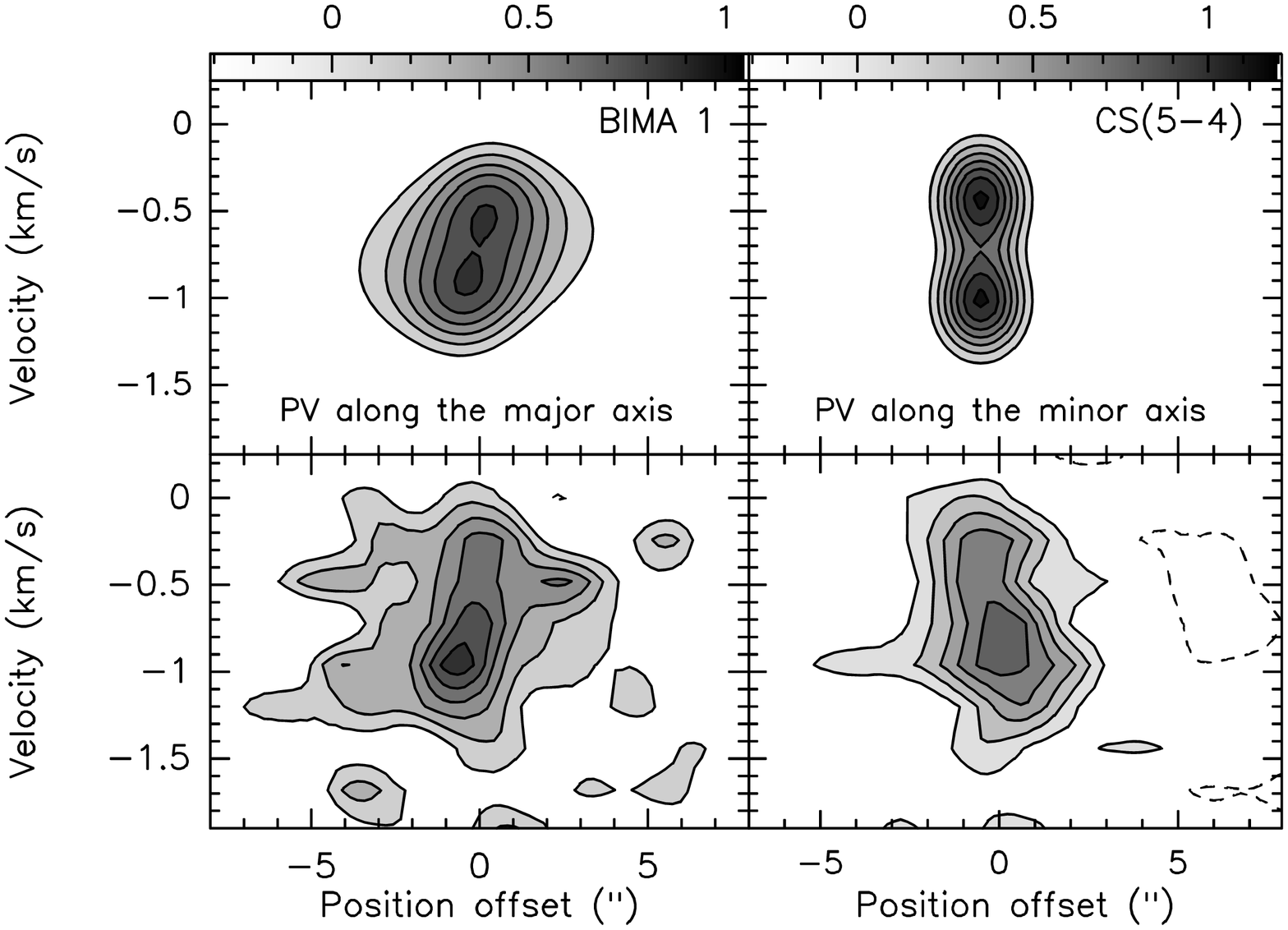}}}
     \hfill
     \caption[]{BIMA~1: ({\it Bottom}) PV plot of the CS \jcc\ emission along the major axis,
P.A.$=110\arcdeg$ {\it(left)}, and  the minor axis, P.A.$=20\arcdeg$ {\it
(right)}, of the BIMA~1 core. The position offset is relative to
$\alpha(\mathrm{J2000})= 21^{\mathrm h} 40^{\mathrm m}  39\fs89$,
$\delta(\mathrm{J2000})= 58\degr 16\arcmin 6\farcs6$. ({\it Top}) Synthetic
emission of the PV plots along the major {\it (left)} and  minor {\it (right)}
axis of a ring with an infall velocity of $0.37\,(R/1000~\mathrm{AU})^{-0.5}$~\kms, a rotation
velocity of 0.15~\kms, and inner and outer radius of 300~AU and
3750~AU. The contours are $-0.3$, 0.3 to 1.05\jy\ {\it (left)} and to 1.20\jy\ {\it (right)} in steps of 0.15\jy.}
     \label{fpvcsbima1}
     \end{center}
     \end{figure} 

     \begin{figure}
     \begin{center}
     \rotatebox{270}{
     \resizebox{9cm}{!}{\includegraphics{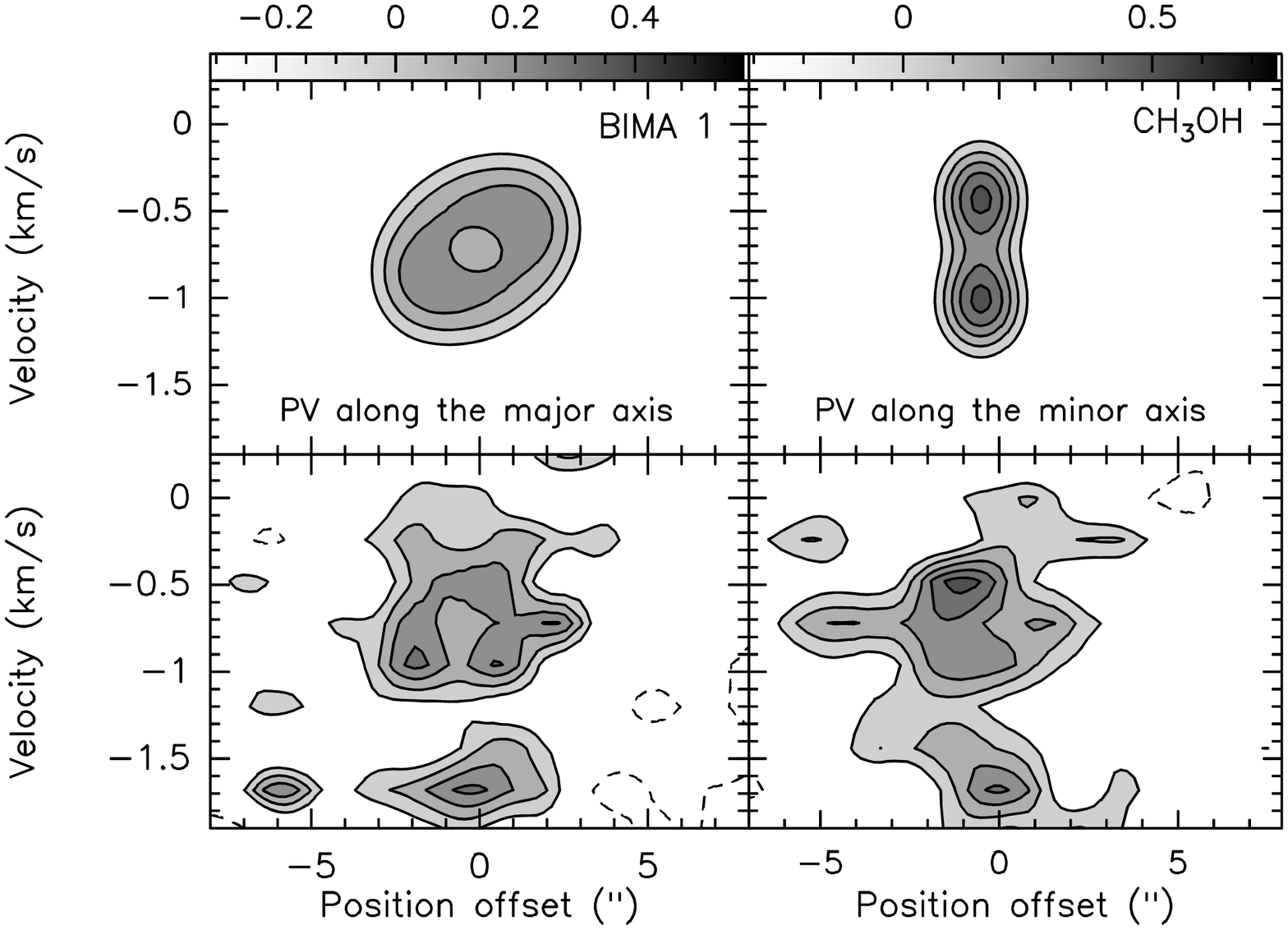}}}
     \hfill
     \caption[]{BIMA~1: Same as Fig.~\ref{fpvcsbima1}, for the \meta\ \jcucu\ emission.
The infalling ring has a radius of 1500~AU. The contours are $-0.24$, 0.24 to 0.60\jy\ {\it (left)} and to 0.72\jy\ {\it (right)} in steps of 0.12\jy.}
     \label{fpvmetabima1}
     \end{center}
     \end{figure}

As seen in Fig.~\ref{fintbima1}, the integrated emission of CS \jcc\ and
\meta\ \jcucu\ shows an elongated core at P.A.$\simeq110\arcdeg$,
perpendicular to the outflow axis with P.A.$\simeq20\arcdeg$. The deconvolved
size of this core is $(5\farcs2\pm0\farcs3)\times(2\farcs0\pm0\farcs06)$, or
$(3900\pm230)\times(1500\pm50)$ AU at the distance of the source. 
Schwartz et al.\
(\cite{schwartz91}) have mapped in C$^{18}$O \juz\ a similar structure toward
the position of BIMA~1. 
If this morphology corresponds to a disk-like structure seen nearly edge-on, 
the ratio between the major and minor axes of the emission gives an
inclination angle of $\sim70\arcdeg$.

The kinematics of the gas surrounding BIMA~1 can be seen in the
position-velocity (PV) cuts done along the major axis 
($110\arcdeg$) and the minor axis ($20\arcdeg$) of the dense core
(see Figs.\ \ref{fpvcsbima1} and \ref{fpvmetabima1}). 
For both molecules, the PV plot along the major axis shows a structure roughly
symmetric both in velocity, respect to a velocity of $-0.7$~\kms, and in
position, respect to the center. The plot along the minor axis shows two blobs
at the central position, symmetric respect to a velocity of $-0.7$~\kms\ as
well. 

For \meta\ \jcucu, the PV plot along the major axis shows a ringlike
morphology, suggesting that the gas is collapsing (or expanding) radially and
that the emission arises from a ringlike or toroidal  structure. The CS \jcc\ PV plot, however, shows a different morphology. The
highest velocities (relative to $-0.7$~\kms) of the core appear only near the
central position, while for the largest position offsets, $\sim\pm4''$, the
relative velocities are smaller. This can be interpreted as the emission of 
superposing rings of different radius, with an infall (or expansion) velocity
that decreases with increasing radius.

 We considered a model of a spatially infinitely thin disk with infall and
rotation, similar to that of Ohashi et al.\ (\cite{ohashi99})  for L1554 and
that of Girart et al.\ (\cite{girart01}) for HH 80N.  Regarding the
inclination, we considered a disk seen edge-on by the observer.  We think that
the inclination $i$ of the disk is small because the PV signature seen in the
data agrees with $i\simeq0\degr$ (see Figs.~\ref{fpvcsbima1} and
\ref{fpvmetabima1}): for both molecules the emission along the major axis
extends roughly from $-5''$ to $+5''$, while along the minor axis it is
compact, and for CH$_3$OH the PV plot along the major axis shows a
characteristic ring shape. In case the disk is not edge-on, one should apply a
correction factor proportional to $\sin^2 i$.  We assumed that the infall
velocity was a power law of the radius, with index $-0.5$ (the value expected
for free fall), $V_\mathrm{inf}\propto R^{-0.5}$.  Since the fit showed that
the rotation velocity was very small,  we assumed it to be independent of
radius. In addition, we assumed that the intensity was proportional to
$R^{-1}$. In case of optically thin emission, this would correspond to a disk
with a surface density proportional to $R^{-1}$. The free parameters of the
model were the inner and outer radii of the disk, $R_\mathrm{inn}$ and
$R_\mathrm{out}$, the infall velocity at a fixed radius (taken to be 1000 AU),
$V_\mathrm{inf}$, and the rotation velocity,  $V_\mathrm{rot}$. The synthetic
PV diagrams were computed along the projected major and minor axes of the disk,
and convolved with a Gaussian in the PV plane, with a resolution of $2''$ in
position and 0.5 km s$^{-1}$ in velocity, to compare with the observed diagrams
for both molecules. 

The observed and synthetic PV diagrams are shown in Figs.~\ref{fpvcsbima1} and
\ref{fpvmetabima1} .  We explored the parameter space of the model and found
that it was possible to found a good fit to the four observed PV diagrams
simultaneously, for the same values of the velocities 
($V_\mathrm{inf}=0.37$~km~s$^{-1}$, $V_\mathrm{rot}\le0.15$~km~s$^{-1}$),  and
different values of the inner and outer radii of the disk for each molecule 
(CS: $R_\mathrm{inn}=300$~AU, $R_\mathrm{out}=3750$~AU; CH$_3$OH:
$R_\mathrm{inn}=R_\mathrm{out}=1500$~AU).

Assuming that the infalling material is in free fall, the dynamical mass
required is given by $M_\mathrm{dyn}=V_\mathrm{inf}^2 R/G$. From the value and uncertainty of the infall
velocity obtained from the fit, we can estimate it to be
$M_\mathrm{dyn}=0.16\pm0.05~M_\odot$. The virial mass 
estimated from the CS \jcc\ intrinsic line width is $M_\mathrm{vir}\simeq0.31~M_{\sun}$ (Table~\ref{tparam})
for a homogeneous density distribution. However, for typical density distributions,  $\rho\propto r^{-p}$, with $p=2.0$--1.5, the virial mass is 0.18--0.24~$M_{\sun}$. Thus, taking into account the uncertainties of
the model, such values are consistent with the dynamical mass. Regarding the
circumstellar mass, the value estimated from the dust emission (Paper I) is
0.07 $M_\odot$, roughly half of the dynamical mass. 
If the observed velocities are infall, this would require a total
(circumstellar plus protostar) mass similar to the dynamical mass, and
thus, the protostar should have a mass similar to the circumstellar
mass. This is consistent with BIMA~1 being a late Class 0 or Class I protostar (Andr\'e et al.\ \cite{andre00}). Thus,
the data available are consistent with the BIMA 1 core being in collapse.

As mentioned above the radii of the ring were $R_\mathrm{inn}=300$~AU and $R_\mathrm{out}=3750$~AU
for CS, and  $R_\mathrm{inn}=R_\mathrm{out}=1500$~AU for \meta.  The value of the outer radius can be affected by sensitivity. Its value is constrained by the extension of the PV map in the position direction. However, this should not affect significantly the estimate of the mass of the disk
because the power-law index for the density obtained is $-1.0$.
Note that the
CS emission spreads over a larger range of radii, up to 3750~AU, while that of
\meta\ is detected only at a radius of 1500~AU. This probably traces molecular
abundance variations with radius  (\meta\ being confined within a narrow range
of radii from the protostar),  or different excitation conditions for both
species. In the latter case the non detection of CS and \meta\ at radii smaller
than $R_\mathrm{inn}$ could be attributed to the fact that higher energy
transitions than those observed are excited for the higher temperature near the
protostar.  Note that a \meta\ knot visible at a velocity of $\sim -1.5$~\kms\
(Fig~\ref{fpvmetabima1}) is not fitted by the model.  The
velocity of this knot does not fit the kinematic structure of the circumstellar
gas. However, the knot may be part of the blueshifted gas, since the blueshifted lobe of the outflow encompasses the position of BIMA~1 (see next section and Fig.\ \ref{fcs21bima1}).

\subsubsection{CS \jdu}
\label{bima1_cs} 

%
     \begin{figure}
     \begin{center}
     \resizebox{9.5cm}{!}{\includegraphics{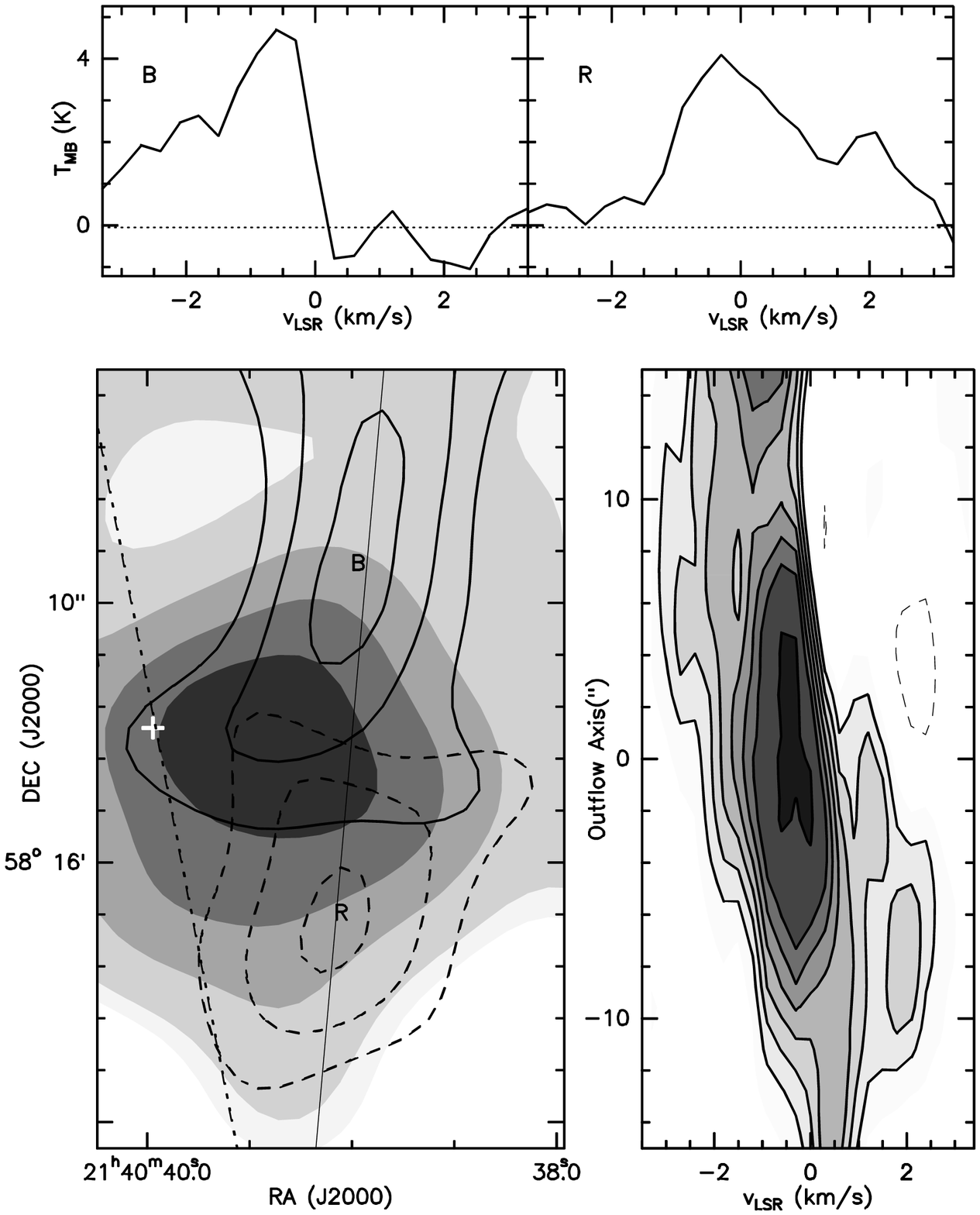}}
     \end{center}
     \hfill
    \caption[]{BIMA~1: ({\it Bottom left}) Overlay of the integrated intensity of the 
CS \jdu\ emission for the velocity interval $(-1.5,0.8)~$\kms\ (core emission,
{\it greyscale}),   for $(-3,-1.5)$~\kms\ (blueshifted emission, {\it thick
solid contours}), and for  $(0.8,3)$~\kms\ (redshifted emission, {\it thick
dashed contours}).  Greyscale levels range from 0 to 1.9\jy\kms. Contours levels are 2, 3, 4, 5, 6, and 7 times 0.15\jy\kms. The white cross shows the 3.1~mm position of BIMA 1 (Paper I).
The dashed-dotted line marks the axis of the north-south CO outflow driven by the YSO
BIMA~1, at P.A.\ $\simeq 20\degr$ (Paper I). The thin solid line marks the axis of
the CS \jdu\ blue and red lobes, with a P.A.\ $\simeq -5\degr$. ({\it Bottom
right}) CS \jdu\ PV diagram along the solid line of left panel. The origin of
the PV diagram is at the position $\alpha(\mathrm{J2000})= 21^{\mathrm h}
40^{\mathrm m}  39\fs00$, $\delta(\mathrm{J2000})= 58\degr 16\arcmin 5\farcs1$. Contour levels are $-2$, 2, 3, 4, 5, 6, 7, and 9 times 0.2\jy.
({\it Top}) Spectra of the CS \jdu\ emission at the centers of the blue and red
lobes.}
     \label{fcs21bima1} 
     \end{figure}
%

CS \jdu\ spectrum in Fig.~2 shows marginally wings of emission, especially at blueshifted velocities with respect to the \vlsr\ of the core,
$-0.7$~\kms. Figure~\ref{fcs21bima1} shows the CS \jdu\ emission integrated for
the velocity intervals corresponding to the dense core, to the blueshifted and
to the redshifted emission. The blueshifted and redshifted lobes are located
roughly symmetrically on both sides of the peak of the CS \jdu\ core emission.
This suggests that the CS \jdu\ emission traces  outflowing gas at moderate
velocities. The axis of this  outflow has a P.A.$\simeq -5\degr$. The PV
diagram along this axis is shown in the right panel of Fig.~\ref{fcs21bima1},
where the blueshifted wing is clearly visible toward the north of the core, and
the redshifted wing toward the south. The spectra of the CS \jdu\ emission
taken at the center of each lobe (see top panel of Fig.~\ref{fcs21bima1})
shows the presence of blueshifted and redshifted  material, confirming the
existence of an outflow traced by the CS \jdu\  emission. Assuming that the CS emission is in LTE and is optically thin, we calculated a mass of roughly 0.02$~M_{\sun}$ for the gas associated with the outflow, for a excitation temperature of 10~K, and a CS abundance of 3$\times 10^{-8}$ (e.g., NGC 1333 IRAS2A: Jorgensen et al.\ \cite{jorgensen04}). The momentum and kinetic energy of the BIMA~1W outflow is $0.05~M_{\sun}$\kms\ and $1\times10^{42}$~ergs, respectively, which are clearly lower than those of the BIMA 1 and BIMA 2 outflows traced by CO (Paper~I). 

The center of symmetry of the CS \jdu\ outflow is located  $\sim 9''$ west of
BIMA~1, a shift that is slightly larger than the synthesized beam diameter
along the right ascension. An inspection of the channel maps at the ambient gas
velocities shows that the peak emission is located between BIMA~1 and the
center of symmetry of the CS \jdu\ outflow.  This is also shown in the
map of the integrated emission for the ambient gas velocities (Fig.~6). All this
suggests that the CS \jdu\ outflow could be powered by a yet undetected YSO,
BIMA~1W, located at the center of symmetry of this outflow (see Fig.~6 caption
for the absolute coordinates). Based on the millimeter continuum
observations (Paper~I), the $3\sigma$ upper limit for the dust mass of this
embedded object would be $\sim 0.04~M_{\sun}$. Note that if BIMA~1W has
associated a compact core traced by CS \jdu, as the other YSOs in this region
do, BIMA~1 and BIMA~1W could not be spatially resolved because of the CS \jdu\ angular resolution,  and the blended emission will appear to peak between
the two sources, as it is the case.

Alternatively, we cannot discard that the CS \jdu\ outflow is not an
independent outflow but it is part of the outflow associated with BIMA~1 and
detected in CO \juz\ at higher velocities (see Paper I). The northern and
southern lobes of this CO outflow are redshifted and blueshifted, respectively,
which is the reverse situation than that of the CS outflow. Yet, this could be
explained if the BIMA~1 outflow axis is near to the plane of the sky. In any
case, in this scenario the CS \jdu\ bipolar emission would possibly be tracing
the denser part of the low velocity cavity of the BIMA~1 outflow.

It is clear that further high angular resolution ($\la 3''$) and high
sensitivity observations at millimeter and submillimeter wavelengths would be
needed to confirm the presence of the BIMA~1W embedded young source and its
association with the CS \jdu\ outflow.


\subsection{BIMA 2}

%
     \begin{figure}
     \begin{center}
     \rotatebox{270}{
     \resizebox{6cm}{!}{\includegraphics{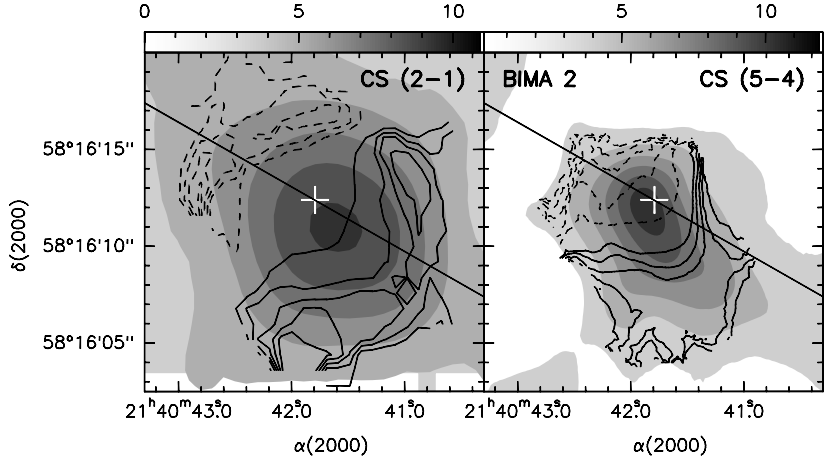}}}
     \end{center}
     \hfill
     \caption[]{BIMA~2: Overlay of the integrated intensity of the CS \jdu\ emission 
     ({\it left}),
     and the CS \jcc\ emission ({\it right}) over the velocity interval $(-3,
     3)$\kms\ ({\it greyscale}), and the velocity field of the dense gas ({\it
     contours}) around BIMA~2. The solid contours depict blueshifted velocities
     with respect to the systemic velocity, and the dashed contours depict
     redshifted velocities. The contours are $-1.5$, $-1.2$, $-0.9$, $-0.6$,
     and $-0.3$\kms\ for blueshifted velocities ({\it solid contours}), and 
     0.3, 0.6, 0.9, 1.2, and 1.5\kms\ for redshifted velocities ({\it
     dashed contours}). Greyscale levels are 0.2 to 10.2\jy\kms\ in steps of 
     2\jy\kms.
     The cross shows the 3.1~mm position of the YSO BIMA
     2. The solid line marks the axis of the east-west CO outflow driven by 
the YSO, at
P.A.\ $\simeq 60\degr$ (Paper I).}
     \label{fmombima2} 
     \end{figure}
%

\begin{figure}
     \begin{center}
     \rotatebox{270}{
     \resizebox{5cm}{!}{\includegraphics{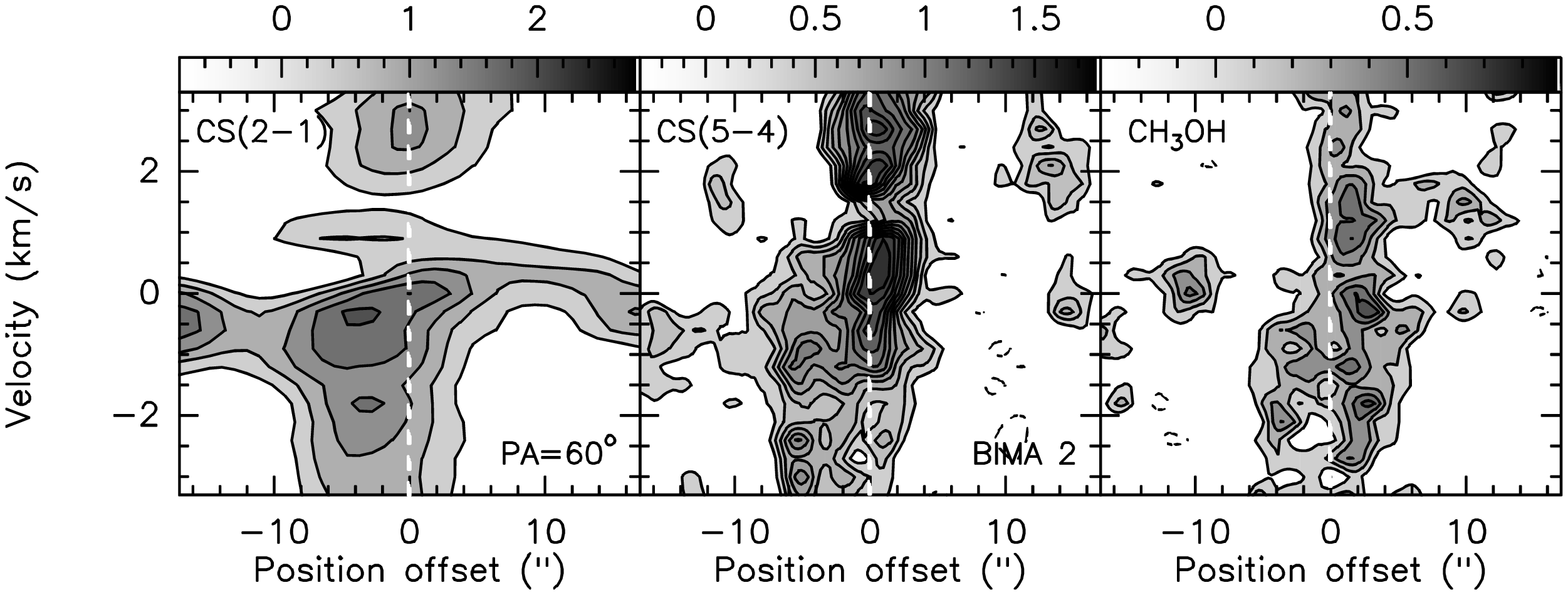}}}
     \end{center}
     \hfill
     \caption[]{BIMA~2: PV plot of CS \jdu, CS \jcc, and \meta\ \jcucu\ 
     along the axis of the molecular outflow driven by the YSO BIMA 2,
     P.A.$\simeq60\arcdeg$, with the offset position relative to the 3.1~mm position of the YSO
     BIMA 2 (Paper I). The contours are $-0.8$, $0.8$ to $2.8$\jy\ in steps
     of 0.4\jy\ for CS \jdu, $-0.3$, 0.3 to 1.8\jy\ in steps of 0.15\jy\ for 
     CS \jcc, and $-0.3$, 0.3 to 0.9\jy\ in steps of 0.15\jy for \meta\ \jcucu.
     }
     \label{fpvbima2} 
     \end{figure}

BIMA~2 is the intermediate-mass protostar associated with IRAS~21391+5802 and
is powering a strong east-west molecular outflow (Paper I). As seen in
Fig.~\ref{fcsto} the molecular core associated with this YSO is the strongest
core in the region. The CS and \meta\ emission are being strongly disturbed by the
outflow even at ambient velocities. Thus, it is not possible to study in much
detail the kinematics of the core because the presence of outflowing gas along
the line-of-sight makes any interpretation of the motions very difficult. This
can be seen in Fig.~\ref{fmombima2}, where the first-order moment (intensity
weighted mean \vlsr) of the  CS \jdu\ and \jcc\ emission around BIMA~2 shows a
velocity gradient across the source in the direction of the molecular outflow
(P.A.$\simeq 60\arcdeg$). This velocity  is correlated with the molecular
outflow; that is, blueshifted velocities toward the southwest and redshifted
velocities toward the northeast. This can also be seen in the PV diagrams of
Fig.~\ref{fpvbima2} that show CS \jdu, CS \jcc, and \meta\ \jcucu\ along the
axis of the molecular outflow. 

%
%

For this core, the CS \jdu\ and CS \jcc\ integrated emission peaks close to the
position of 3.1~mm continuum source (Paper I). However, Fig.~\ref{fpvbima2}
shows that around BIMA 2 the CS \jdu\ emission peaks at $\sim-0.5$~\kms\ 
whereas the CS \jcc\ emission peaks at $\sim$0.5~\kms.  This along with the
clear self-absorption feature seen in both transitions at the position of the
dust source (Fig.~2) is likely due to a high optical depth of
ambient gas emission.  Therefore, this suggests that most of the CS emission at ambient
velocities is not tracing the dense core around BIMA 2 but the interaction layer
of the molecular outflow with the dense core. This is supported by the
kinematical correlation of the CS emission at ambient velocities with the
molecular outflow.

\begin{figure}
     \begin{center}
     \rotatebox{270}{
     \resizebox{6cm}{!}{\includegraphics{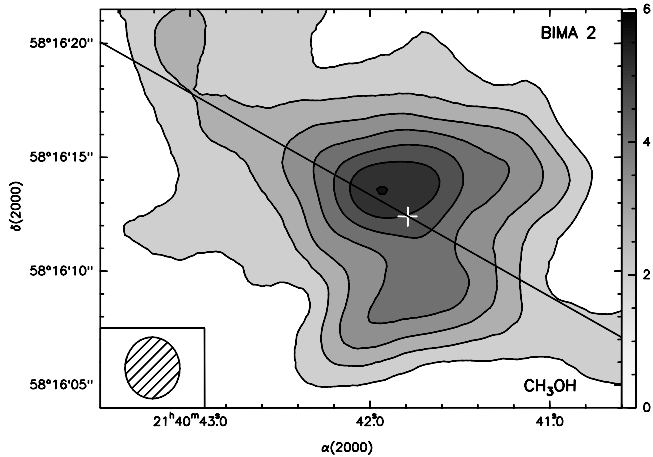}}}
     \end{center}
     \hfill
     \caption[]{BIMA~2: Integrated intensity of the \meta\ \jcucu\ emission toward BIMA~2. 
     The cross shows the 3.1~mm position of the YSO BIMA 2. Greyscale levels 
     are 0 to 6\jy\kms\ in steps of
     1\jy\kms.  
     Solid line marks the axis of the east-west CO outflow driven by 
the YSO BIMA~2, at
P.A.\ $\simeq 60\degr$ (Paper I). The synthesized beam is shown in the lower 
left-hand corner.}
     \label{fch3ohbima2} 
     \end{figure}

The \meta\ \jcucu\ emission peak is offset by $\sim 2''$ with respect to the CS
\jcc\ emission peak, of the order of the beam size, in the outflow direction
(see Fig.~\ref{fch3ohbima2}).  The PV cuts of Fig.~\ref{fpvbima2} show a
deficit of \meta\ emission toward the position of the 3.1~mm source for all
velocities. This molecule is usually produced by low-velocity shocks induced by
the molecular outflow and it is more often found in association with the
outflowing gas (e.g.\ Bachiller et al.\ \cite{bachiller98}). Thus, this
suggests that the chemistry of \meta\ is probably being affected by shocks
between the molecular outflow driven by BIMA~2 and the dense core at ambient
velocities. However, the spatial offset between the CS and \meta\ suggests that
they trace different outflowing material, due either to different chemical
conditions or to different excitation conditions.

In summary, for this intermediate-mass protostar, both CS and \meta\ are not
good tracers of the dense core at ambient velocities but of the interaction of
the molecular outflow with the core. This scenario is different from the
low-mass case, where the high-density gas surrounding the protostar is clearly
tracing the dense core at ambient velocities. Thus, in order to study the kinematics of the dense core associated with BIMA~2, observations of different high-density tracers are needed.

%
%

\subsection{BIMA 3}
\label{bima3_diss}

%
     \begin{figure}
     \begin{center}
     \rotatebox{270}{
     \resizebox{7cm}{!}{\includegraphics{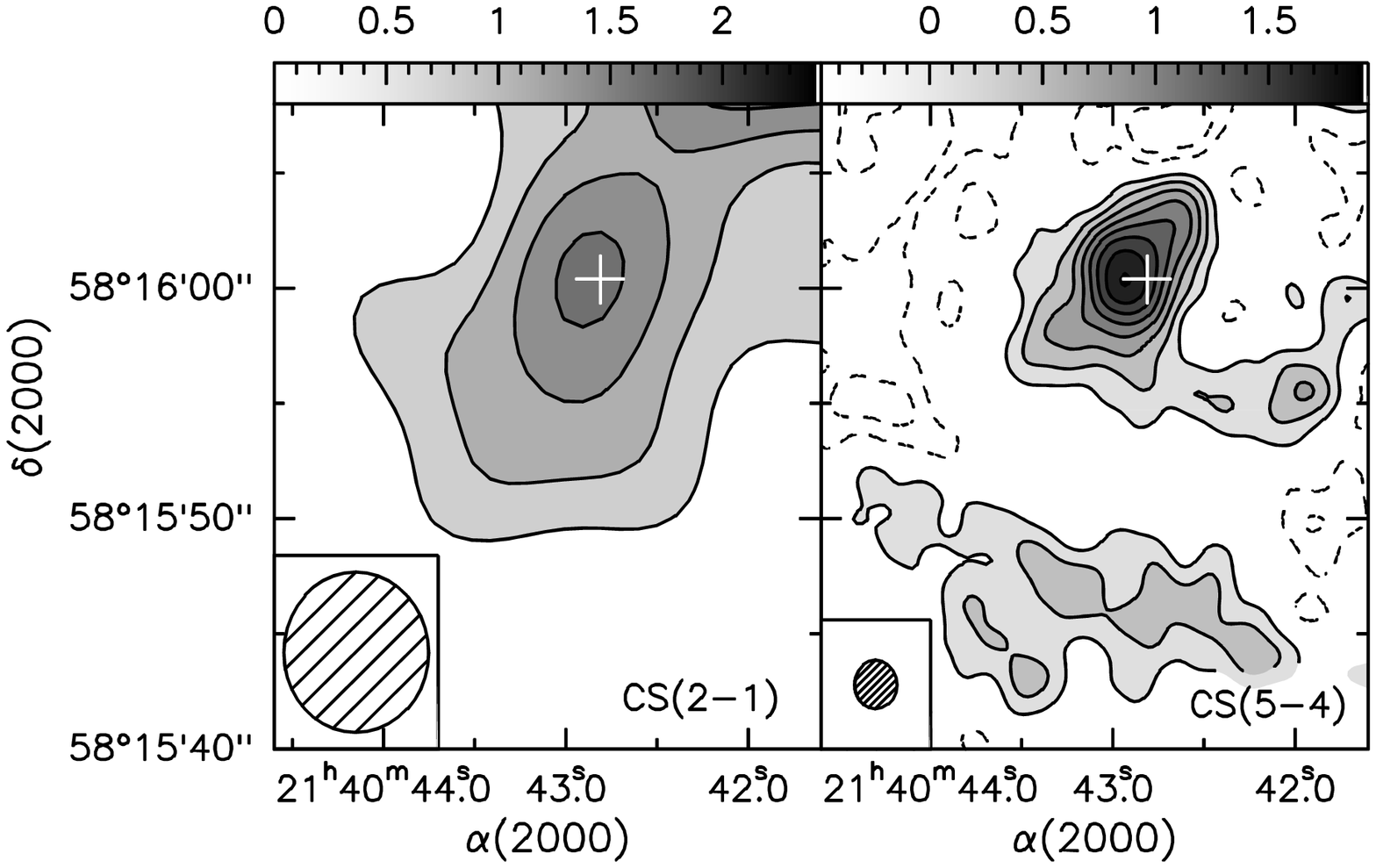}}}
     \end{center}
     \hfill
     \caption[]{BIMA~3: Integrated intensity of the CS \jdu\ and CS \jcc\ emission over the velocity interval $(0.5, 3.0)$\kms\ toward BIMA~3.
The contours are 0.56 to 2.24\jy\kms\ in steps of 0.56\jy\kms\ for CS \jdu, and $-0.48$,
$-0.24$, 0.24 to 1.92\jy\kms\ in steps of 0.24\jy\kms\ for CS \jcc. The cross shows the
position of the 3.1~mm source BIMA 3 (Paper I). The synthesized beams are shown in the lower left-hand corner.}
     \label{fintbima3} 
     \end{figure}
%

The molecular core associated with the YSO BIMA~3 has been clearly detected in
CS \jdu\ and CS \jcc, and marginally, in \meta\ \jcucu\  (see Figs.~2 and
\ref{fintbima3}). However, no molecular outflow was found in association with
it in Paper I. This core is well defined but it is clearly at a velocity
different to that of the BIMA~1 and BIMA~2 cores, as shown by the spectra
(Fig.~2) and by the different moment maps of the velocity field. The \vlsr\ of this
core is around $\sim2$ \kms. The CS \jcc\ integrated emission shows a
northwest-southeast flattened core, which has a deconvolved size of
$4\farcs1\pm0\farcs1\times2\farcs9\pm0\farcs04$, or
$3100\pm75\times2200\pm30$~AU at the distance of the source, at a
P.A.$\simeq150\arcdeg$ (see Fig.~\ref{fintbima3}). 

\begin{figure}
     \begin{center}
     \rotatebox{270}{
     \resizebox{5.8cm}{!}{\includegraphics{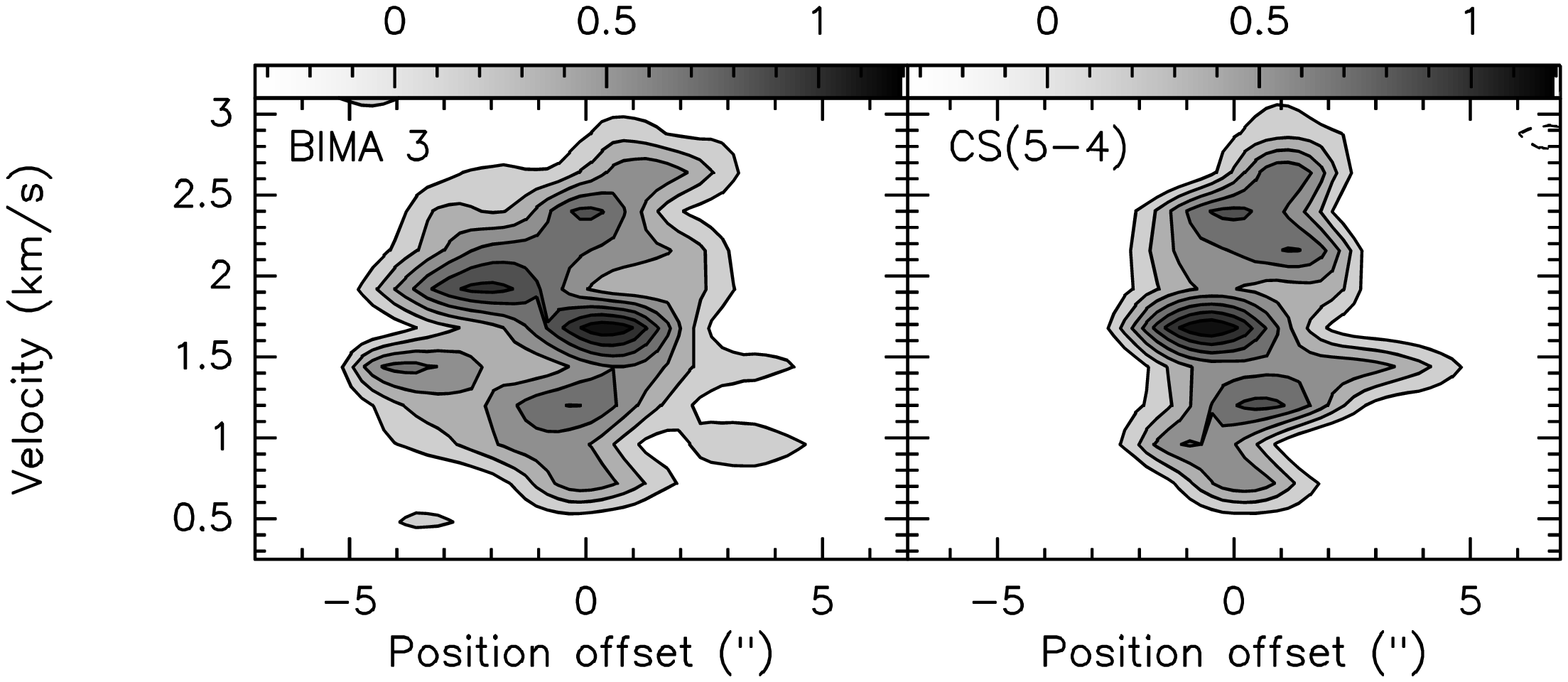}}}
     \end{center}
     \hfill
     \caption[]{BIMA~3: PV plot of the CS \jcc\ emission along the major axis,
P.A.$=150\arcdeg$ {\it(left)} , and  the minor axis, P.A.$=60\arcdeg$, {\it
(right)} of the BIMA~3 core. The position offset is relative to
$\alpha(\rm{J2000})= 21^{\rm h}40^{\rm m}42\fs91$, 
$\delta(\rm{J2000})= 58\degr16\arcmin00\farcs3$. The contours are $-0.3$, 0.3 to 1.2\jy\ in steps of 0.15\jy.}
     \label{fpvbima3} 
     \end{figure}

CS \jcc\ PV cuts done roughly in the direction of this elongation
(P.A.$\simeq150\arcdeg$) and in the direction perpendicular
(P.A.$\simeq60\arcdeg$) are shown in Figure~\ref{fpvbima3}. These diagrams are
similar to those of BIMA~1 (Fig.~\ref{fpvcsbima1}), and thus, they could be
indicative of collapse or expansion of the molecular core. The CS \jcc\ PV cuts could not be
properly fitted by our simple model (the \meta\ emission is too marginal to
include it in the fit). This suggests a non-uniform density distribution of the circumstellar material around BIMA 3, which might be due to fragmentation and the presence of subcores within BIMA 3.  Alternatively, the infall process in BIMA~3 could be weaker than in BIMA~1.


%
%

\subsection{Evolutionary stage of the dense cores}

A detailed study of the emission of the  molecular cores at ambient velocities
surrounding the intermediate-mass protostar IRAS 21391+5802 has allowed us to
infer or confirm  the evolutionary stage of the cores studied in Paper I.

For YSO BIMA 1, we suggest a model for the CS and \meta\ emission, proposing
that it is a low-mass object undergoing collapse. From the CO~\juz\ data
(Paper~I) we estimated that the dynamical timescale for the north-south
molecular outflow centered at the position of BIMA~1 is of the order of 10$^4$
years. Such a value is consistent with dynamical timescales estimated for
outflows driven by Class~0 objects (e.g.\ Andr\'e et al.\ \cite{andre00}; Richer et al.\ \cite{richer00}). Near BIMA~1 the detection of
a CS \jdu\  bipolar outflow suggests the existence of a yet undetected YSO,
BIMA~1W. The dynamical timescale estimated for the CS \jdu\  outflow is
1--2$\times 10^4$~years, which suggests that the powering object is also a very
YSO. However, further observations are required in order to confirm this new
source and derive its properties.

We found that the CS and \meta\ emission associated with YSO BIMA 2 is strongly
disturbed by the outflow, indicating that the protostar is in a very active
stage of mass loss, which confirms that this intermediate-mass object is a very
young stellar object, probably a Class~0  object. The dynamical timescale estimated for the CO~\juz\ east-west molecular outflow driven by BIMA~2 (Paper~I) is of the order of 10$^4$~years, which also confirms that BIMA~2 is a very YSO.

Regarding YSO BIMA~3, it is much less massive ($M\simeq0.07\,M_\odot$; Paper~I)
than BIMA~2  ($M\simeq5.1\,M_\odot$; Paper~I), and its emission is more
compact.  This core is associated with maser emission (Tofani et
al.~\cite{tofani95}) and centimeter continuum emission (Paper~I). Both are
clear signposts of star formation and indicate that the outflowing phase has
already started. Furthermore, the detection of \meta, although marginal,
indicates the occurrence of shocks in this core, likely produced by an outflow,
although can also indicate mantle evaporation due to the heating from the
protostar. However, although those observations suggest that BIMA~3 is a YSO, 
no large scale molecular outflow emission has been detected, and no clear and conclusive infall evidence is visible from the PV plots (see Fig.~\ref{fpvbima3}). Thus, both infall and outflow are probably weaker for BIMA~3 than for BIMA~1, which suggests that BIMA~3 would be a more evolved
object, possibly a low-mass Class~I object. The molecular outflows powered by
such objects are usually much weaker than the ones driven by Class~0 objects (e.g.\ Bontemps et al.\ \cite{bontemps96}), which would explain the non detection of the molecular outflow toward BIMA~3.


None of the YSOs BIMA~1, BIMA~2, or BIMA~3 has been detected at $J$, $H$, or
$K_s$ band through the infrared 2MASS survey. However, in the immediate
vicinity of the YSOs there are other infrared (proto)stellar sources detected
at $K_s$ band ($2.17~\mu$m). This means that in a relatively small region there
are very young stellar objects, detected at millimeter wavelengths (the BIMA
sources) together with more evolved young objects (the 2MASS sources).

\section{Conclusions}

We conducted a kinematical study and modeled the gas emission of the molecular
cores at ambient velocities surrounding IRAS~21391+5802, an intermediate-mass
protostar embedded in the core of IC~1396N. The high-density gas has been found in association with three dense cores associated with the YSOs BIMA~1, BIMA~2, and BIMA~3. 

For the YSO BIMA~1, the integrated emission of CS \jcc\ and \meta\ \jcucu\
shows an elongated core at a  P.A.\ of $\simeq110\arcdeg$, perpendicular to the
outflow axis. We modeled the gas emission by considering a
spatially infinitely thin ring with infall and rotation seen edge-on by the observer, with the intensity and infall velocity distributions being power laws of the radius.
The best fit was obtained with the intensity proportional to $R^{-1}$, an infall
velocity $V_\mathrm{inf}=0.37\,(R/1000~\mathrm{AU})^{-0.5}$~\kms,
and a rotation velocity $V_\mathrm{rot} \leq 0.15$~\kms.  Assuming that the
infalling material is in free fall, the dynamical mass required is
0.16~$M_{\sun}$, which is similar to the total (circumstellar + protostar) mass
in the core. 

The CS \jdu\ emission toward BIMA~1 peaks at a position $\sim5''$ west from
that of the 3.1~mm continuum, CS \jcc, and \meta\ emissions. The high-velocity
CS \jdu\ emission is tracing outflowing gas at moderate velocities ($\sim
2$~\kms). The center of symmetry of this CS outflow is near the CS \jdu\ 
emission peak. This outflow could be powered by a yet undetected YSO, BIMA~1W,
or alternatively could be part of the BIMA~1 molecular outflow.

The CS and \meta\ emission associated with the intermediate-mass protostar
BIMA~2 are not tracing the dense core but they are highly perturbed by the
bipolar outflow even at velocities close to the systemic value. This confirms
that the  protostar is in a very active stage of mass loss, and also supports
the idea that intermediate-mass outflows interact more dramatically with the
dense gas surrounding the protostar than those driven by low-mass protostars. It also suggests that CS and \meta\ are likely not 
good tracers for studying the properties of dense cores associated with intermediate-mass protostars.

The core associated with BIMA~3 is at a systemic velocity different to that of
the BIMA~1 and BIMA~2 cores. The CS \jcc\ integrated emission shows a
northwest-southeast flattened core.  The lack of molecular outflow and of
clear evidence of infall suggests that both outflow and infall are weaker than
in BIMA 1. Thus, BIMA 3 is probably a more evolved, Class I, object.

CS \jcc\ PV cuts are similar to those of BIMA~1, and
thus, they could also be indicative of collapse of the molecular core, with an infall velocity of $\sim0.4$~\kms\ at
a radius of 1000~AU. This infall velocity corresponds to a dynamical mass of
0.14~$M_{\sun}$, comparable to the total mass.

\begin{acknowledgements}
     
 We thank the referee Debra Shepherd for her valuable comments. RE
acknowledges support from Generalitat de Catalunya grant 2002BEAI400032. RE,
JMG, MTB are partially supported by MCyT (Spain) grant AYA2002-00205.

This research has made use of the NASA/IPAC Infrared Science Archive, which is
operated by the Jet Propulsion Laboratory, California Institute of Technology,
under contract with the National Aeronautics and Space Administration. 

\end{acknowledgements}

{}

\end{document}